# Holistic Natural Systems – Design & Steering
## Guiding New Science for Transformation

Jessie Henshaw, HDS natural systems design science

sy@synapse9.com

Contents   17,200 words

## Abstract (447 words)

Great societies and their cultures, like all natural systems (and as opposed to conceptual systems), emerge from their environments, organized and behaving as wholes with their internal designs coupled with their external worlds. So nature can be a great teacher of what complex holistic designs are and how they successfully work. Some notably become imperiled by challenges of their own making, as ours has; driven to endlessly maximize its growth naturally leading to ever-growing conflicts, internally and with nature. Now we can compare different kinds of growth systems in their natural contexts to expose their different ways of coupling with their contexts and steering. Some work out fine – the difference often how internal and external parts fit. It helps identify how some take paths leading to deep trouble while others work out fine – the difference is often with how internal and external parts do or do not fit together. Better steering (self-control) also comes from more exposure to internal and external contexts, allowing more prompt notice of new situations and a sure response.

Recognizing emerging systems starts with noticing something new becoming a growing center of relationships, a nucleus of activity in a nourishing place, something sprouting. Storms, trees, people, businesses, organizations, cultures, etc., all start as emerging internal designs that build themselves using connections with nourishing contexts. That coupling between internal and external worlds continues to evolve as the new system makes its *home in the world,* lasting for a short or long time. Another coupling of internal and external worlds matters, too, between human thoughts and lives. Our mental worlds are only indirectly connected with our contexts and can blind us to the meanings of life, as in one of our earliest recorded experiences, not feeling at home in the world. Those feelings of alienation, doubt, and separation from nature, turn out to greatly affect how we design our living systems, even helping to make reality only seem conceptual.

So with this "kit of parts," we explore emerging system steering using familiar examples. A simple diagram asks good leading questions to remind readers what emerging system designs and non-verbal cues for response to notice for successful steering. We also read the meaningful *progressions* as arcs of stories about relationships. We first get perspective from multiple views, like noticing smooth or rough takeoffs and landings as cues to look all around before guessing how they happen. Their likely validity comes from confirming their *nonlinear continuities* of *emerging design*, which are hard to fake, making the stories one reads into them *reasonable hypotheses* to check out. Finally, we use ordinary language to refer to natural systems in context, not abstractly, using careful language as our first systems science.

Electronic supplementary material

| | |
|---|---|
| Manuscript | http://synapse9.com/_ISSS-22/MS-HNS1-Design&Steering.pdf |
| Talk slide set: | http://synapse9.com/_ISSS-22/Talk-HNS1-Design&Steering.pdf |
| Workshop slides: | http://synapse9.com/_ISSS-22/WS-HNSI-NoticingSystems.pdf |

Credit Author Statement: Funding, Concepts, and Figures by the author UON.

Keywords: emergence, whole systems, growth, germ, seed, coupling with contexts, organizational stages, innovation, individuation, inflection, turn forward, coordination, release, engagement, steering, conceptual blinders, resolving complexity, holistic senses,

# 1 Introduction

*"Nature is always more subtle, more intricate, more elegant than what we are able to imagine." – Carl Sagan[1]*

*"If we knew what it was we were doing, it would not be called research, would it?" – Albert Einstein[1]*

In a scientific study of natural systems, the first question has to be, "What can one seem to know for sure when it is so clear we cannot know very much?" Nature is more than complex; it is everywhere, independently organized and *animated*[2]. No wonder our world is so confused by what we ended up gowing into, a world led by brilliantly educated cultures unable to collaborate in managing an increasingly unmanageable world seemingly headed for near-term demise. The one place we seem to see hope is that near-term demise has seemed likely at nearly every turn of our 200+ history of rapid exponential growth, only with more reason now. The demise seemed to be pushed off in the past, only to return by creating and using some new form of social and economic organization. Can we do that again?

It does pay to look at the great array of dead ends we seem to be facing (Henshaw, 2020).[3] One seemly remote chance at present is for people to follow nature's tried and true formula for resolving growth crises for long life. It would involve changing what the system invests its profits in (money or resources), from maximizing growth to supporting the emerging new life's effort to learn how to cope with its challenging new world. That would mean our choosing to use profits to care for what profits built, the innovative systems we need to live, and make them sustainable within their means. That is like what graduates and other hatchlings, new businesses, and organizations go through when forced to end their implausible experiments and make things work.

---

[1] Science Quotes – http://www.planetofsuccess.com/blog/2019/science-quotes/

[2] Italics is used for key terms that refer to natural physical circumstances structures and processes

[3] An experimental list of World Crises Growing with Growth - https://synapse9.com/_r3ref/100CrisesTable.pdf

### The natural systems approach

That posture is what we take here, not only because natural systems are complex in so many ways and engage with others. It is also because so many seem more controlled from inside than out, like people and the weather. Each identifiable individual also seems to emerge from within the larger whole too. So it is impractical and likely misleading to devise "controlled experiments." So finding what one *can* know requires extra clarity in observation and language. There is some precedent for this as a scientific method (Boulding, 1953, Goethe, 1996). Here we focus on reading the evolution of emerging system designs associated with their growth, as is what ultimately seems to power natural systems.

One of the initial questions is whether natural systems *have* organization, or are they just coincidences of uncertainty? The latter is what science has implied by representing how nature works with the statistical implications of recorded data. Whether natural systems are forms of organization or not can be looked at in many ways. It is also a great test of one's perception to see what makes the difference. One good start is to look up the etymology *organization*[4] which roughly means "made like an organ," and from a systems view, "working as a system."

I call it "new science" for natural systems due to the focus on studying how natural systems develop their own rules, freely experimenting within the bounds of the fixed laws of physics. One can follow how that happens by watching systems grow to become self-defined and self-governed, coupled with an external environment; like us, developing by emergence of new design, not *cause-and-effect*. In general, cause-and-effect implies the subjects are out of context, and *organization principles* are not considered, such as the emergent properties of new relationships or of connecting complementary parts that apply to things in context (Bateson, 2017; Henshaw, 2008).

---

[4] Etymology Online - https://www.etymonline.com/search?q=organization

Here we also start by recognizing that long ago, humans developed language as their first systems science. It was our way of recording important patterns, relationships, situations, experiences, identities, etc., and attaching our feelings and other meanings to share. A pair of the earliest now common words from the dawn of language are mother[5] and father.[6] They name roles in a family and society and identify the radical (at the time) new social unit and organization of the nuclear family. When that occurred is unclear, as it may have been before there were words for it. That two-syllable pair, *mo-ther,* and *fa-ther,* seem present in nearly every ancient language, though, along with the related *ma-ma* and *pa-pa*. There is a dispute about whether these nearly universal terms are shared from one language to another or come from the earliest sounds infants naturally make, which they do seem to.

They also correspond to the nuclear design of family life, an invention that may have spread like the words from culture to culture.[7] So, we can use common terms both to refer to some of our most ancient and durable knowledge of life and the *self-defining* things, arrangements, and features of life with which we live and work. That way of defining terms is very attractive as a source of terms for natural systems for scientific study, as the whole systems they define and other methods of scientific definition are unable to do. It turns out that the natural systems of life are far better defined by themselves than any observer will ever be able to, with the possible exception of the fundamental laws of physics, amazingly useful generalizations of things too small to observe, with one exception. That is the implication that the laws of physics are universally deterministic, disputed by the organizations of the parts that seem to develop on every scale of the known universe, and the bounty of emergent properties of new kinds of natural systems also display (Volk, 2017, 2020 ).

## Observation

We see natural systems behaving as a whole, as "things holding together" such as water drops to the solar system, various organisms, and other forms of organization. How they work and what holds them together are mysterious for many reasons. We generally cannot either see inside them from the outside or see the whole from any place inside. So we often build an image that combines what we see, often not with uncertainties and questions about what to follow. Another problem is that a whole system does not work by "cause and effect" so much as by stimulus and response, engaging networks of relationships. So, all parts tend to be responsive to every other. Like a family, a system's internal designs develop individually to be self-sufficient except for access to its coupled networks of outside sensors and connections for information and resources. How they work is a little more exposed as their designs and connections develop during their growth, so that is one of the focuses of study.

---

[5] Mother https://www.etymonline.com/search?q=mother

[6] https://www.etymonline.com/search?q=father

[7] Question of ancient aboriginal Australian use of the terms, for example
https://www.westernsydney.edu.au/dhrg/digital_humanities/featured/past_projects/mama_and_papa_in_indigenous_australia

Exactly why systems form by growth is a little mysterious, starting imperceptibly small generally, in some enabling context, from some initial *germ* or *seed*. So, to an observer, they always begin when first noticed. Also complicating the identification is that growth starts so slowly, but regularly, adding new parts and building up activity, going from less to more systematic by taking faster-and bigger steps, a nonlinear and *compound growth* pattern. As a result, novices may not have preconceptions and notice them well before experts who constantly compare what they see with what they think they know. That is why experts, poets, and others learn how to return to being naïve observers at will. It is to have perceptions that are more open and truthful.

Why nature starts all systems with compound growth was the subject of the previous paper on the nature of natural systems (Henshaw, 2021). How growth starts is a kind of explosion, which makes it quite visible as a sign of emerging systems of system change once one gets accustomed to seeing at a cue to notice what is happening. However, we can never have more than limited information on what is happening. We can often only see when systems noticeably change, but awareness of contexts and periodically paying close attention help. So it is getting to know them and watching for cues to respond, as much as we do habitually in raising children or doing projects at work, that expands our vision of what is happening in the world that matters to us. Natural systems science is about watching for cues in familiar and unfamiliar contexts and learning to convert perceptions into information for others using familiar language.

> Reflection – *What it comes down to is that we do not define reality.*
>
> That is what nature does! We just observe and take simple notes on things that are quite complex we call "concepts." That, unfortunately, detaches our notes from the contexts that would have given them meaning. So to build rather than lose meaning, we need to use concepts to help see how much they leave out as an ongoing learning process. It works best when making observations to pause at the end to get a broad sense of the contexts, as well as taking away the facts. That makes it much easier to discover new things when you go back for what else there is.

One successful way to study the organization of systems is to look for patterns from one place that also occur in others, say patterns of individual character, delayed reaction, overreaction, etc. That validates some questions while raising more, forming a tree of branching learning. For example, it can help to compare kinds of boundaries and look at what relationships do and do not cross the boundary. With that open approach to revisiting the same question finding new information both enriches and validates that the inquiry is of natural rather than conceptual designs. When checking theories, one must look for exceptions whenever the theory seems repeatedly confirmed. Nature will always give only similar answers.

The usefulness of theories and concepts varies considerably. A more advanced but quite practical method is to first develop a theoretical model and then use it, by contrast, to aid in searching for how nature departs from it. Theories are always abstractions (Back, 2006) that simplify and detach a natural pattern from its context but can also refer to the natural phenomena and contexts of interest. For example, when increasing pressure on a system, the result may be unexpected, such as causing it to *jitter* before it fails. That exposes another level of organization and raises a broader question about how systems under pressure will unexpectedly behave. Disturbing pressures can have horrible effects on people, such as causing many strange kinds of panic, while in other circumstances causing beautiful overtones from musical instruments. However, that pressure can cause unpredictable disturbance also tells us that all systems have internal designs that can be disturbed. That offers a very useful cue to respond to systems behaving unexpectedly, to see what may be pushing the limits of their resilience. How to use these techniques will be one of our focuses here.

As we study signs of development and change, we see more evidence that systems are not numerical in design but organizational, built by forming networks of relationships that create bonds by working together. However, numerical variables of interest are still useful for prompting us to look to nature for answers, asking the main question: "What are we seeing, and can we check?" For that, we generally need to accumulate contextual information as evidence and search and digest our perceptions using a combination of *reasoning* and *feeling*. That makes full use of our wonderfully and essential but underrated holistic senses. Our feelings help to enrich and balance our reasoning, and reasoning to enrich and balance our feelings. It does depend on being open one what one finds, but that process of enriching and validating one's perceptions can ground one's thinking deeper and deeper in reality.

For example, the question "is the bolt tight" seems like a simple question. However, it turns attention to both the bolt and its contexts and what might have prompted the question, like concern about the cause or dangers of a loose bolt. Asking why the question came up is where the variables are, prompting a search of the context. So, in general, expecting to learn something new every time you ask a question about a natural system is a great sign that one is learning from nature rather than looking for abstract answers. If not, one might be asking the question wrong or perhaps asking it of a model or theory that always gives the same answer. What does that mean? It means that any meaningful information from nature will reflect *all* the relationships in at least the immediate context.

## 1.1 The fascination

The fascination with how nature works by itself, what this study came from, did not come from trying to imitate nature but from noticing all the little transients that seem to begin and end all events. They turned out to be the *takeoffs* and *landings* of systems emerging with compound growth. It came up in freshman physics during a demonstration of using a strobe light to trace a digital path of a ball in gravity, producing a simple parabola one can then calculate. So, to make a small joke, I asked, what about the tossing and catching? I thought it funny that we only discussed the part of the behavior that closely followed a fixed rule. The brief dynamic beginnings and ends of things generally do not.

Later I found that split attention between animated and deterministic subjects was rather global, with most of the sciences looking only for fixed rules. Perhaps it was just having not found useful ways of studying emergent systems with individual designs and behaviors everywhere in nature. It might also have been that science has been unusually effective in making things profitable from its beginning. Hence, as a culture and often also doing work funded by profits, for methods of control, it would have grown faster than other forms of inquiry and tend to dominate. Also, naturally funded from profits, it would be natural for funding sources to ask it to study how to control and be certain of profits, steering the culture of science toward assuring the most profitable methods of control. In any case, I left physics and went on to study architecture, where how complex designs emerged and came to completion, the beginning and ending transients of lasting services, is always very important.

> <u>Reflection</u> – *The cost of our simply tremendous history of success*
>
> For centuries science and business mined profitable rules of cause and effect. That seemed to come with a solitary focus for developed societies on growing our control of nature and growing money, now quite visibly having distorted all of human consciousness. Perhaps that is what blinds everyone to the nature of emerging systems and the other amazing secrets of nature's success while neglecting our tremendous emerging mortal threats of its overshoot!

## Comparing growth systems

The method introduced here for comparing different kinds of natural systems in context is a new kind of natural systems science. There are limitless variations in how systems grow, most having identifiable unique individually. That comes from growth being an *individual*, *internal*, *animated*, and *exploratory* process that occurs in an also individually unique context. To study them, we start by comparing three generic internal models and three internal design strategies for growth (Figure 1). Then, we study and compare the shapes of their development curves as records of *system learning,* looking for patterns of internal and contextual design and development and potential cues for responding to them. That is also how we compare natural systems with those developed by people, generally finding that they have different structures but similar growth processes.

We look for how an emerging system's interior and exterior worlds are coupled, how each navigates its challenges and otherwise behaves. We study natural self-organizing systems as self-animated since what allows them to develop is a design for capturing energy, but still not having human characteristics. So "half alive" and often taking off in unexpected directions, like whirlpools, air currents, and fire, exhibiting behaviors we also see in people and social networks. Systems displaying them include astronomical nebulae (Wiki-a, 2022)[8]. Even single-cell organisms display elaborate forms of such behaviors, like slime mold. Naturally occurring systems that include people and human-made systems that include people will display human characteristics but still not be human. So, it is important to use clear language.

What can you say about the systems that develop along the growth paths seen in Figure 1. They could be either human social systems or biological or chemical systems. Those kinds all produce different growth curves like those seen here. Are the shapes associated with different animated processes? Are the ways the curves change characteristics of any familiar internal or external conditions likely to affect a system's behavior?

---

[8] Astronomical nebulae https://en.wikipedia.org/wiki/Nebula

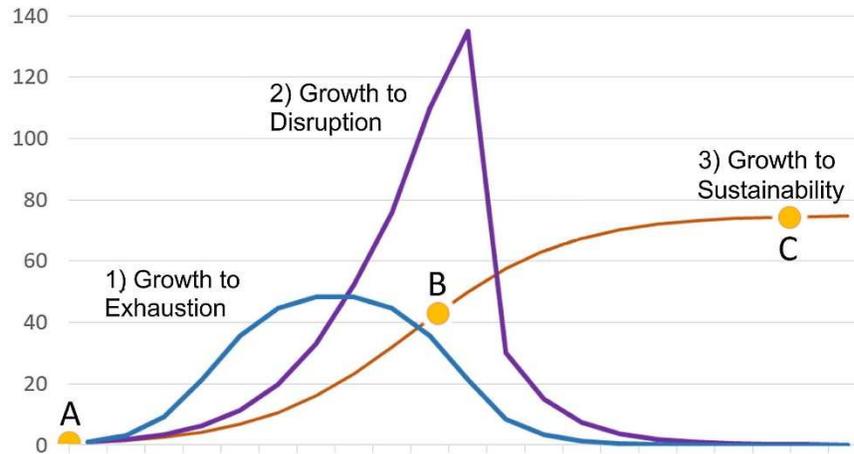

Figure 1. <u>Degrees of endurance for emerging growth systems</u>: Failing for being uncreative, Failing for being too creative, and Having it just right.

All systems appear to require successful germination by some new combination of things that can concentrate energy and resources for building up their animating structures, creating a *viral* or *contagious* process. For systems that show Type 1) Growth to Exhaustion, resources run out, starting with growth as the system forms by using resources to build itself, but then declines following some kind of bell curve. For Type 2) Growth to Disruption systems, the resource is relatively infinite, or the system's creativity lets find other resources. Without any constraint, growth might continue to accelerate until it disrupts its internal processes. That case lets us compare a match that pops and blows itself out without consuming all the phosphorous with creative human relationships, businesses, and societies with boom-to-bust system designs. That opens up many examples to study with much the same problem. How many familiar examples can you think of for Type 1) and 2) growth? The one uses a resource till it runs out. The other expands faster and faster till it upsets its internal workings.

Then some systems grow and avoid those growth hazards, Type 3), accomplishing all three transitions, A, B, and C, to have a relatively long period of sustained climax before meeting some later decline or hazard. That kind of system must be internally creatively adaptive enough to change with its circumstances. It needs to expand beyond the limits of its initial resource and then shift to using a sustainable resource, no longer using its internal building capability to expand exponentially but to maintain the health of its internal design and environment. That means having the capacity to sense cues to avoid hazards, i.e., developing an instinct for self-preservation and survival. There are various examples of simple energy or chemical systems that are resilient and self-regulating as if having a survival instinct. However, that is not what we mean to focus on here. A simple case of a near-living system adapting to its environment to become sustained for a while was that of a warm air current. It was temporarily blocked by another and then, as if "waiting its turn," resumed its travel on its original path. It is not important but a reminder that non-living systems also sometimes creatively adapt.

So to apply the comparison of those three growth system types, we can ask:

- Which of the three paths seems most like the world's we live in?
- How are we doing at reading the cues for changing our strategies?
- Civilizations can live quite long, and ecologies and species much longer. If we are so inventive and perceptive, what are we missing in the world are we missing.
- The people running the planet tend to be the best educated and most communicative types of people from good families; where is their mistake?

### Basics of systems steering

Passive environmental influences such as resources, pathways, and constraints, to which natural systems make an animated response, show systems navigating their terrain using *internal system steering.* That would apply to all organisms, of course. It would perhaps also apply, if differently, to weather systems, which are internally animated and respond to external change. But, on the other hand, one would not call it system steering; if a physician, gardener, or other knowledgeable caregiver gives *support* or *assistance* to other things. That is *system guidance* instead, not *steering* unless you consider the guiding person a part of it; then, it is a system of mutual steering of the guide and the subject. Externally *forced steering* of a system not as care would be *interference*.

| | |
|---|---|
| Caring for a home and garden | Taking a trip with others |
| Continuous self-correction, such as like homeostasis | Following a practiced script or method |
| Responding to threats and opportunities | Hunting for things matching or complementing a given pattern |
| Responding to external cues to take action | Responding to internal cues to take action |

Table 1. <u>Different kinds of system-steering</u> for people or other systems with internal guidance help them navigate. Think about the varied purposes, conditions, and cues for response required for each.

> Ostrom – Guides to common interests. (1990, 2009)
> Exposing regional stakeholders to their contexts helps them understand what is happening and make better choices.
>
> Midgley – Community Involvement. (2007, 2011)
> Exposing community stakeholders to their contexts to help them understand what is happening and make better choices.
>
> Cabrera – DSRP systems thinking. (2008, 2022)
> Deepening perception of reality by looking for how four dimensions of reality connect: Distinctions, Systems, Relationships, and Perspectives,
>
> Smith – AIC systems management. (2009, 2013)
> Expanding the powers of Appreciation, Influence, and Control from and on internal, transactional, and contextual domains.
>
> Henshaw – Growth as natural system design. (2018, 2021)
> Connecting how we and nature both make new systems by expanding on expanding germinal forms that start and change scale as wholes.

Table 2.<u>System thinking methods for enhancing the learning experience</u>

The self-steering of natural systems is the subject here, how animals, communities, or other *self-preserving* systems are alerted and respond to threats and opportunities in their paths. We mostly aim here to broaden the context of thinking around the similarities and differences between the strategies for steering ourselves and steering our planet. Both have challenges to overcome and opportunities to grasp.

Figures 2 & 3 below show pictures of the *defensible domains* of a family home and a car on a road trip as examples and symbols of the more general relation between a natural system's *internal* organizational center and its *transactional* and wider *environmental* contexts.

The practical steps of steering such systems in all variations follow the same general pattern. There are four separate or blended steps: *sensing*, *responding*, *preparing*, and *acting*. For these two examples of the home and a traveler, the specifics of any sensing and response for one of either kind could be quite different. We know that intuitively, but it is also because *steering* is an extended series of *emergent designs*.

Examples of merging the four steps into a single flowing process are important. They often include steering cars, coordinated body movement, dance, flocking and conversation, and of course, the self-governance of economies. Hesitant response; pausing to recalibrate with each response is also common. For example, birds often move that way, as do artists and designers who move one careful step at a time. Then there are people pressed to respond but unable to, who freeze instead, their "animal spirits" kept from moving them into action at all until they recover. Another interesting pattern is strategic hesitation, such as birds waiting before taking flight or people waiting to initiate a plan, waiting for the right conditions.

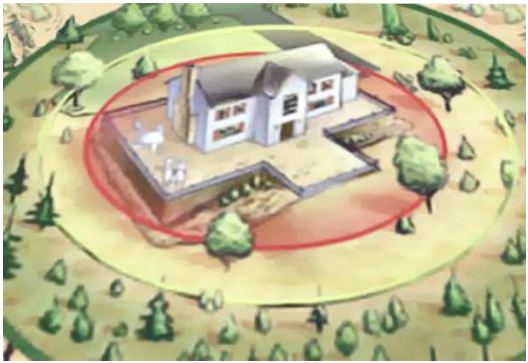 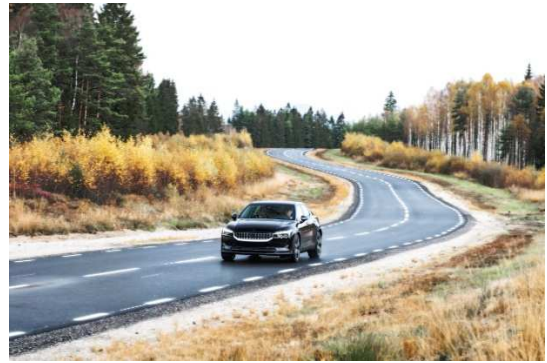

Figure 2.   Defensible domains #1             Figure 3.   Defensible domains #2

Suggesting the nesting of the *Internal*, *Transactional*, and *Contextual* domains of living systems for engaging its powers of (or corresponding to) *Reasoning* and *Feeling* for navigating environments, responding to a contexts *Distinctions, Systems, Relationships,* and *Perspectives*

Steering, then, seems to rely on feeling as a primary guide. You would not think that, given how people say they are reasoning their choices of what to do. How many kinds of feelings we have is not clear, but very many, and they are essential for our survival instincts and help us feel our way along the steering demands listed in Table 1. For feeling and action to be in sync and steering choices to flow smoothly, sensing and response coordination mechanisms must be on the levels of both autonomic nerves and whole system behavior. We certainly react and feel that when responding to an unexpected threat and also while cruising along expertly steering a car while talking or thinking about other things. You might say feeling one's reasoning and reasoning one's feelings is a mind-body collaboration central to steering.

**Unchangeable whole systems**

The evidence that many kinds of natural systems begin, develop, work, live, and only change as wholes includes how our bodies grew from single cells. Our growth in the womb changed our form a few times, and after we were born, our bodies and lives also changed form repeatedly while retaining the distributed individuality of our characters and behaviors. Part of what makes enduring individual wholes possible is the coupling of the self-contained internal system and its connections with stable external systems – a coupling of a separate world with connections. That is also what makes every enduring family in its home so unique and enduring; little can interfere with its defensible internal world or its connections. It is similar for cultures and languages developed by a growth process as wholes, but without a physical skin or enclosure, as "open systems.". Their organizational structures make them unique and able to be flexible and adaptable as whole systems, but only able to change as wholes. The primary "force" holding them together is the emergence of versatile properties from making complementary connections. Simple examples are having internally controlled openings between *interior and exterior spaces* or having *cars on streets*. Even simpler examples include *knife and fork* and *pen and paper*. It genuinely seems that everything *made* is made of these.

Cultures and languages can only change the structures that make them whole, their *tensegrity*, by internal processes *of the whole* to change what holds them together. The key to their combined durability and flexibility is that they have something like a "blockchain design," with "authentic copies" of the design for the whole system like a genome, distributed to every part for active use or reference. The specific means of whole system coordination are often more visible than how a genome unifies an organism. Even in that case, we can't follow how the whole system coordination works. We do see there must be a coordination mechanism, though, and see direct evidence. That validates looking for more.

For each cell to check if what other cells make fits the master plan is also something a body would likely need to survive. So together, that gives one a puzzle for helping connect related observations. For familiar cultures and languages, the puzzle is how they remain so self-consistent over such long periods and are also so flexible and innovative. We seem to see the coordination mechanism in every cultural or language engagement, relying on two things. One is distributing authentic copies of the master design by acculturating people to that way of living or speaking, in youth or by immersion later. That gives them "the authenticated copy" to check against others for consistency, "the distributed ledger." For languages and cultures, the authentic copies contain a large part of their history, much like every family conversation relies on having access to much of the history of family conversations.

If one watches closely, each connection with someone starts with checking to see if there's enough recognition as being part of the same culture or having enough shared roots of meaning to understand one another. Computers don't understand, but we design them to repeatedly authenticate all transmissions. How else could you have confidence about what is being "transmitted." So in cultural or language relationships, we rely on connecting with the roots to share meanings, which maintains the root meanings forever since the root meanings originally come from root experiences with others and nature. That's the real foundation. The one way the root meanings change, which does happen, is by innovation in meaning that extends the culture or language as a whole.

So it seems that the most successful of humanity's inspired, proud and ancient cultures found their way into such a troubled world as we find today, perhaps by having taken a wrong turn some time ago, while all along evolving to work and stick together as a whole. So like all of us individually, our cultures can only repair our path by changing as a whole. Of course, we'd still need to keep checking the new and old root meanings as we go, but to relieve our troubled steering of the planet travel to safety, it seems necessary for it to result from inspirations of the whole.

---

### Physics and the science of natural systems

Both physics and natural systems science aim to offer reliably useful abstract models of how nature works. However, their use is quite different. The use of the one is predictive and of the other exploratory. Natural systems science is about what we can discover about the organization of nature, something that numbers cannot represent. There is also a disagreement in some quarters about whether nature works the same as our most useful abstract models. Physicists seem mostly non-committal, simply relying on the equations that fit the data for predicting controlled or uncontrolled outcomes and represent the universe with grids of numbers.

In contrast, natural systems science uses models to assist with discovering forms of organization in nature, prompting targeted questions about continuities that depart from the simpler paths of models. For example, a car will never follow the exact shape of a road, but that shape is a good place to start when looking for the true path that a car did or would likely take. For example, that might help identify which skid marks were associated with a particular car's path of travel.

For another difference, physics assumes "cause and effect," implying that there will be no effect lacking an external force and that models of external forces determine events. The models based on it have proven exceptionally powerful but do not raise questions about the contexts in which they are used. So our social systems have gotten into serious trouble, unaware of the consequences of using them.

The premise of natural systems science is different. It is to look for and study *emerging individual systems,* their causations, internal designs and relationships, evolution, life cycles, self-steering, and their external connections and relationships (Henshaw, 2015, 2018, 2020, 2021). The real beginning of this approach to systems science was discovering a new way to use the central law of physics, the conservation of energy. It was to ask: How can physical systems begin and end while maintaining continuity of their energy needs and uses? The answer that turned up was "emergent organization" achieved by the deeply nested internal physical processes of what we see from the outside as flowing numerical "growth" (Henshaw, 1995, 2010).

Not all kinds of natural systems are like those that develop as wholes through an organizational building process we call *growth*. However, they are the most useful kinds of natural systems that people are part of, work with, work on, or work through. They are also special because they are designs of nature powered by developing complex organization more than by quantities of energy. So, they can regularly break some of the "laws of nature" for a while, like some of the laws of thermodynamics. So, one might ask if this stuff called "natural system organization" is something real or imaginary, a new kind of proposed "dark matter," maybe?? How else might it have slipped by the attention of our many highly advanced fields of modern science?

The terminology comes from what the words of natural language refer to in nature, so by definition to what is real (not theory), and what the inventors of language saw and asked about as natural phenomena. Western languages all get the term *organ* more or less intact from the ancient root Indo-European terms '*werg*' and '*ano*' that mean respectively 'work' and 'do'[9] Combined, they mean "thing that works." So in the word '*org-an*' the prefix '*org-*' refers to *the thing,* and the suffix *'-an'* to the *doing*, and "*org-an-iz-ations*" can be read as "*creations made of working relationships.*" So natural systems, as the "*working relationships created for the work of nature*," seem to be both organizations and organs in the traditional sense. So, no dark matter, just a better understanding of the meanings of the common language.

It is such an important subject that one would expect it to be well studied. It seems that the lack of study may be due to the reality that natural processes define having a form that cannot be represented in numbers or theories, such as people define. However, better questions about natural forms can be raised using numbers and theories to illuminate natural designs if used well for that purpose. That poses a problem for physics, which can only approximate nature with random variables in any case. A further barrier to representing natural systems with physics is that they emerge from their environments by internal causation, composed of complex systems of relationships complexly coupled with environmental context. So though this approach to studying the working relationships that let nature's systems work by themselves has only begun to develop, the progress relies on standards related to physics and has come up with some useful new questions.

---

[9] Organ – Etymonline: https://www.etymonline.com/search?q=organ

## 1.2 Ancient Cultural Roots

*"The concept of alienation identifies a distinct kind of psychological or social ill; namely, one involving a problematic separation between a self and other that properly belong together." – David Leopold*

**Conceptual blinders**

There are two fundamentally different ways of learning. One is to absorb direct impressions from senses, observations, and experiences without interpretation, as we all do at times, especially at unguarded moments. Artists and skillful observers do it for their work and others for connecting deeply with reality. But, of course, that would have been the only way humans learned at first and what began to change as we developed language. It is unclear when humans started to rework our impressions of life to replace them with more powerful abstract concepts for life. It was probably in stages before or after our brain's neocortex developed, so a range between 90,000 and 50,000 years ago seems plausible (Wayman, 2012).

Concepts are powerful because we can arrange and change them quickly, giving thought an ultimate modeling tool and toy for rearranging ideas of reality. They can work wonderfully in combination with our accumulated direct observations, feelings, and impressions, each informing the other. However, because concepts are inventions and easy to manipulate when we first learned to build entire realities from them, it makes individuals and communities vulnerable to becoming quite confused. The problem arises when making decisions based only on concepts when they either blind you to or you simply lack any accumulation of direct observations, feelings, and impressions of the reality your decisions will impact. This potentially systemic loss of connection with reality does seem to be the source of, perhaps not all, of course, but certainly many of humanity's major mistakes, like humanity's current accelerating destruction of our home on Earth. The evidence that a blind fixation on a concept is the cause of that disaster is that we have long had a rule for endless compound growth. All of nature uses compound growth only as a brief kickstart for building systems.

There is extensive evidence that humans have become confused about our relationship with reality. First, there is the very long history and innumerable kinds of alienation we experience.

Alienation[10] is fear of or despair over losing attachment with the world, other people, or other communities. It is both a common modern and ancient experience. The story of Adam and Eve seems to record it as Adam eating from the "forbidden fruit of knowledge" and as the "fall of man" (i.e., alienated from God). What more perfect poetic description of humankind becoming lost in its own false realities could be? There are also the wars, all the wars, which rely on soldiers alienating the people they face in battle — another seemingly near-perfect match with concepts resulting in a loss of context.

There is also the strange matter of worldwide cooperation on maximizing the compound growth rate of our consuming and disrupting the natural world. Perhaps the oddest part of it is how the people managing the world catastrophe are the elite of the educated class. They are the leaders in high-paying positions in the world's governments and education, research, business, and institutional communities. They are simply the best of us, well liked, well mannered, and responsive to others in every way at home and in their communities.

The problem seems to be with the stark black and white difference between the manipulation of concepts we are driven by at work and the enjoyment contexts at home. Even sustainability was redefined as "business as usual" by a little manipulation of concepts to make business work. See footnote below for links to further context.[11]

Ah, to feel at home in the world. It happens now and then; some people sustain it for long periods, of course, then also usually have to rejoin society again to "get along." Many people have a natural ability to easily form, attach, and detach from their conceptual worlds. A home is virtually always a place for being real. The battle between home life and global power run amuck will not be won by finding a way to be happy oneself. We need more from us to make the world happy it seems.

---

[10] (Leopold 2018). *The Stanford Encyclopedia of Philosophy* https://plato.stanford.edu/entries/alienation/

[11] Further author research context in Reading Nature's Signals: https://synapse9.com/signals
  Institutional malfeasance - How Sustainability became BAU
  https://synapse9.com/signals/2022/02/18/how-sustainability-became-bau/
  Why we see life as conceptual - Betrayed by the power of our minds
  https://synapse9.com/signals/2022/03/13/betrayed-by-the-power-of-our-minds/
  List of ever-expanding crises - The Top 100+ World Crises Growing with Growth
  https://synapse9.com/_r3ref/100CrisesTable.pdf

## A Long-Lived Hestian Culture

Great societies and their cultures grow from small beginnings, usually beginning somewhere a local culture of competence developed. Greece and Rome are familiar, and many other strong societies have developed too. Some lasted far longer than we expect societies to last today, like long-lived bronze age societies.

Conceptual thinking for systematically using technology to build cities began about 10,000 years ago (Whelan, 2020). When that combined with law, finance, accounting, and central government, allowing leaders to collect profits to use in increasing their power, creating a boom of boom and bust civilizations, seems marked by the many short-lived Bronze age Mesopotamian city-states (Grossman & Paulette, 2020; Wyse & Winkleman Eds., 1999; Van de Mieroop, 1997)). However, the slightly older Egyptian, Minoan, and Aegean island civilizations were different, more interested in the arts of living than power. These seem to be the source of our modern world's mixed heritage of art for art's sake and power for power's sake (Burkert, 1985; Dinsmoor & Anderson, 1973; Henshaw, 2015).

The bronze age is dated roughly from 3000 BCE to the beginning of the Greek Dark Age, 400 year period between the collapse of the Mycenaean civilization, around 1200 BCE. That was followed by the emergence of the Greek Archaic Period, around c. 800 BCE, coinciding with when Homer's work appears to be the first use of writing to record vivid stories. The best-known of the long enduring civilizations of the Bronze age are those of England (Pearson, 2009), the Aegean (Burkert, 1985) (Dinsmoor & Anderson, 1973), Minoan Crete (Willetts, 1977), Egypt (Wilson, 2013), and China (Loewe & Shaughnessy Eds., 1999).

According to Burket as well as Dinsmoor & Anderson, a long-lived proto-Greek Aegean culture lasted about 2000 years, traceable by its unusual ritual home design, centered on a low hearth called the *Hestia*. The Hestia was at the center of a large gathering space wide enough to need internal columns. The Hestia was where the home kept its perpetual flame but was usable for warming food too, so with surrounding space for gathering, the home was well designed for long large meetings. The archeological evidence places examples of that design from the beginning to the end of the Bronze age. It is found at the bottom layer of Troy in 3000 BCE and in the Minoan and Mycenae palaces just before the Greek dark age.

Its design even became the model for the revolutionary innovations of Greek architecture, with classical Greek temples copying the form. The design is also still associated with Hestia, called and perhaps literally the "*the first of the gods*" in the role of "*guardian of the sacred flame of hearth and home,*" the low hearth for the perpetual flame of the home still called "the Hestia," too. That remarkably long enduring central institution of an advanced pre-Greek home culture also led to classical Greek architecture, democracy, arts, and sciences. Modern western traditions of hearth & home are also directly inherited from its devotion to the sacred life of the home (Dinsmoor & Anderson, 1973). Curiously that very long-lived and historically important culture seems to have no name, so we can call it *the Hestian culture*.[12]

## Then Growth for Its Own Sake

*"The warlike states of antiquity, Greece, Macedonia, and Rome, educated a race of*

*soldiers; exercised their bodies, disciplined their courage, multiplied their forces by*

*regular evolutions…" – Gibbon*

*– The Rise and Fall of the Roman Empire –*

We also inherited from ancient Greece its later classical traditions of the public sphere (Polis[13]), the name for the city states of classical Greece as administrative and religious urban population centers that later became centers of wealth and power (Egen, 2004). So, it represented "new culture" in historical terms, for the building of Greece's city cultures and a counter force to the much older Greek home sphere (Oikos) that inherited Hestian culture.[14] Part of that change was due to the 350-year interceding dark age between the long establishment of Bronze age Hestian culture and the emergence of classical Greek culture, some 150 years later. With the latter also came the emergence of Greek science and the quick discovery it could be extremely profitable (Engen, 2004; Farrington, 2016). Thales[15] was the first scientist, a gregarious Ionian trader who sailed the Mediterranean on business, also collecting mathematical principles from every ancient culture. Though the records are scarce, he first gained fame using his maths to make a fortune in the olive market, as if by inventing futures trading. That seed of how to design systems for making piles of money would not have disappeared, of course, though there seems much less about it in the records.

---

[12] One of the more fascinating features of the well documented ancient archeological and cultural heritage closely associated with what may be the actual first of the Greek gods, with a 2000 year tradition, having great influence even today on Greek culture, is the near total absense of reference to it in mainstream histories. It is as if mainstream histories were restricted to stories of city, wealth and and army builders, which the Hestian culture, as advanced as it was in many ways, was not.

[13] *Polis* the Greek city-state, or "public sphere" as opposed to the private Oikos, https://en.wikipedia.org/wiki/Polis

[14] O*ikos* is Greek for the unity of the family, its property, and its home, https://en.wikipedia.org/wiki/Oikos

[15] References to Thales' Science and Philosophy https://www.google.com/search?q=thales+science+philosophy

After discovering how to use math to turn a small sum into a large one, with little effort, the people who learned to use it to grow their power also left many more records of what their power did than how they did it. They built much bigger and more successful societies with advanced technology than the boom and bust societies of the past. Their great success was partly due to rewarding their populations rather than exploiting them. It would have tempted everyone to join in on the limitless boom headed for its natural bust – for using power to multiply power – ending in internal, external, and environmental crises and conflicts. It is the same formula now followed by modern society and its world economic culture. A much more complete story of cultures that built economies designed to fail is the book by Joe Tainter (1988), the Collapse of Complex Societies. His general assessment may well be the most insightful. He concluded that they all seemed forced to create solutions for their problems that were too complicated, seeing increasing "complexity" as the killer it certainly is. So, for example, depleting resources requires more complex efforts to obtain them, resulting in declining resource availability and a society requiring increased resources to operate. It is called EROI, energy return on energy invested (Hall, Balogh, & Murphy, 2009; Henshaw, 2011; Lambert et al., 2014).

The clear evidence is that in the shift from home-centered to national cultures, as ancient cultures transitioned into modern ones to grow profits using technology, some of the most basic principles of life were left behind. The cultural knowledge of how to live developed by the relatively advanced Bronze Age home cultures held on for centuries. It is also still with us in our own home cultures, now about 3-5,000 years from when it developed. That is partly because societal and home cultures have never mixed well, preserving the ancient ways. Our senses of individuality and alienation (Leopold, 2018).

The initial germ of urban design seems to have been 10,000 years ago with the technology of organized farming to sustain settlements of traders and artists as non-farmers. Our recent and still accelerating explosion of urban life began quietly with the Renaissance when the world economy began doubling in scale every ~350 years (Maddison, 2008). That continued until "great acceleration" (Steffen et al., 2015), which began upon Watt's perfection of his rotary steam engine in 1780. That abrupt start of our now threatening global explosion is most visible in the data on global atmospheric $CO_2$[16] (Henshaw, 2019), also showing our long history of multiplying CO2 as fast as we could be continuing. Since 1780, the world economy has been doubling at nominally every 25 years, almost ten times, so increasing in scale by about 1024 times.

---

[16] The Scripps record of combined icecore and atmospheric CO2 (scroll down to the figure).
https://scrippsco2.ucsd.edu/data/atmospheric_co2/icecore_merged_products.html

## 2  The Form of Natural Systems

**Rigidity and flexibility**

One of the most important and fascinating general features of natural systems is their common combination of very stable structures that can only evolve as a whole, with highly adaptive parts simultaneously. Take our bodies, a rigid structural design with very adaptive parts. Our structural designs capable of evolving are mostly our ways of life and thinking, often with groups of people adopting new ways of thinking in new situations. For example, when faced with an emergency, people tend to shift to thinking only about the new common threat very quickly. There are also the many counterexamples of human thinking becoming notoriously rigid. There are so many ways. We get stuck on habits, rules, theories, cultural and social customs, and strategies that work in some places, but we seem stuck with them everywhere. It is what Gestalt psychology called "functional fixedness"(Wiki-b, 2022)[17] A natural systems view and its power to ground us to reality can help fight these sometimes quite dangerous possessions.

We are also aware of personal feelings of alienation produced by mental or sometimes physical barriers we and others erect, preventing personal, professional, or cultural connections. It is a common feeling that these unwanted separations seem to oddly throw everything we want to be secure into doubt. Overcoming barriers of self-isolation to keep others out also often wonderfully enhance community engagement. But unfortunately, self-constructed barriers now keep our world from focusing on its many current common existential threats. It is not just the existential threats of climate change but the vanishing of the natural species and environment, the congestion and confusion and other increasing pressures on human societies, and many more (Henshaw, 2020). Given this pattern of highly abnormal widespread misbehaviors, our *polycrisis,* some call it. We should comb our experiences for similarly converging multitudes of differing crises, "plagues of plagues," on any scale or at any time to learn from it. It is rather common once you understand what you look for, such as things *going haywire* in multiple ways due to too disturbing internal pressures. We should even consider the ancient oral traditions telling of similar systemic distress, such as the ancient stories of the Bible.

---

[17] Functional Fixedness - Wikipedia https://en.wikipedia.org/wiki/Functional_fixedness

The above outline of a new exploratory way of learning from repeatedly consulting natural examples of related kinds is a scientific method, unlike the traditional one in some ways but quite compatible as an addition to it once understood. Like traditional science, it searches for meaningful patterns of relationships to then validate and build on. What's different is focusing on complex natural systems of relationships, not abstract representations. Focusing on watching for cues to respond to from systems and their contexts becomes both a method of deeper investigation and one of system *steering* instead of *control*.

### The lifecycle of natural systems

Science is about finding what one can seem to know for sure when it is also clear we cannot know very much. For example, the conclusion that natural systems develop from tiny starting designs comes partly from recognizing the many common terms we have for what initiates larger scales of organization, such as *nucleus*, *egg, seed, spark, eye, kernel, spirit, germ, stem, urge, inspiration, notion,* or *idea*.[18] We call what those initiate: *growth*, *sprouting*, *development*, *propagation*, *crystalization*, *germination, animation,* or the *emergence* of *systems*, *relationships*, *roles*, *work*, *play*, etc.[19] We are also unable to find exceptions to systems developing from discrete but tiny beginnings. We do find development processes and generally find a burst of non-linear self-organization associated with their beginnings, and then either good evidence of some minute seed to start them or do not find anything.

That also seems implied by general physics as an implication of energy conservation. The implication is that new energy uses need to develop by a continuous succession of increasing scales of energy using processes, which we do generally observe, to maintain the continuity of energy use required by energy conservation (Henshaw, 2010). Science relies so heavily on nature, exposing her work for us to study; it may be hard to accept that the sources of new forms of natural organization are generally impossible to observe. An easily understood case is the formation of snowflakes. We see that an ice crystal forms at the beginning, but not how its complex design blossoms from it. It does indeed suggest some kind of order is present at the particle/quantum level of matter, but we may also never have a way to observe it due to the uncertainty principle.

---

[18] There are also a very wide variety of names for *ideas* that begin things, like *animus* and *amity*
[19] The etymology of *inspiration* as: *in-spir-ation*, gives it is original meaning as things receiving the breath or spirit of life, [Genesis ii.7].

The start of a new system's organizational development is generally also in a quiet place with available energy and other resources. The germination first captures energy to invest in building up ways to capture more, forming a driver of positive feedback for multiplying organization that begins the system's growth. At the same time, the system's connections with its environment grow developmentally, too. That gives the system as a whole the form of a coupling of its emergent internal design with its external networks of contextual relationships, a new "*life*" emerging from its environment. That starts the period of *syntropy* for the system, creating *new organization* and *concentrating energy* coupled with *entropy* for the environment. The result is like a tree with roots, which in the tree's case has *environmental roots* both in the ground and the air.

The tree also has an unusual extended syntropic life, continuing to build its biomass and concentrate energy until it is old and stops growing. Human lives are something like that, too, continuing to expand and concentrate resources and influence until near the end of their lives. These patterns may vary quite a bit, of course. What is constant is the usefulness of asking the questions raised by the normative life cycle of living systems (Miller, 1973), Figure 4, which shows the normal case for systems that endure beyond their initial burst of formative growth. They grow syntropically until reaching a peak of vitality and resilience as they mature, then maintain syntropic processes as they entropically age while enjoying a long period of environmental engagement before declining. That typical cycle for new system lives seems to fit them all; if series of stepwise emergence, engagement, and decline stages are included.

All the stages would be accumulative and have nested scales and stages of development at one scale that create environments for the next. Most often, the observable transitions from one state to another follow fairly smooth S curves, as if the development stages and the whole and the parts are all composed of processes of regular proportional change. Why small scale progressions create, large smooth shapes may fall to the conservation of energy, that every scale of change needs to develop without discontinuity. It might also come from other benefits of regularity.

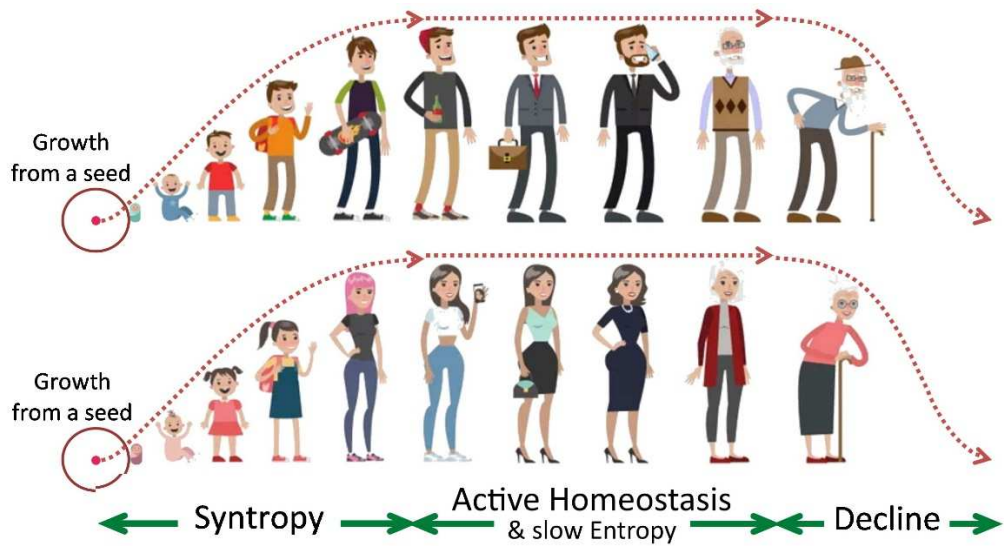

Figure 4. The normative physical life cycle for living systems: Following their initial A) explosive growth from a seed and then a longer period of B) maturing and learning from their new world, then C) long period of environmental engagement sufficient for reproduction – the stages of sustainable life. The scales and stages of development vary for individuals, species, and cultures: drawings – Guzaliia Filimonova/Getty images.

In nature, every system begins with compound growth (Henshaw, 2021). It is a process of using small amounts of emergent power to multiply power. That is quite essential for systems to develop from their tiny initial scales. However, that process could also destroy the originally sustainable design if continued until it pushed the system beyond its organizational upper bounds. The signs of pushing systems beyond their limits are something we are all experienced in responding to if noticed early enough, and it is quite common in caring for normal personal and workplace projects and relationships. Those are some things that "make life what happens when we were planning something else." The ability to respond to exceptions is what makes systems naturally tolerant and responsive to variation. By their nature, systems work as wholes and *stand together*,[20] having reserve capacities and organizational resilience as basic properties of their organization.

For healthy emerging systems, like the growth of human lives in the womb, *birth* occurs as a response to the growth limits of the womb and the need for experience to make it in the world. The mature fetus is fully formed as a human but highly undeveloped and faces the major environmental change shock of being thrust into a strange new environment. That seems to be nature's optimal solution, starting things up and hoping a shocking new challenge will prompt them to take hold. In some ways, that seems to fit our current world situation, too. Our world culture is both fully formed and shockingly undeveloped, stuck with having pushed the earth's limits to cause the organization of our systems to begin breaking down while severely limiting our ability to respond. It will no doubt be inconvenient. It looks like it is time to find some other version of nature's plan for shepherding new lives to enduring success.

---

[20] The original root meaning of *sys-tem*. https://www.etymonline.com/search?q=system

A simple model for raising the detailed questions we would need to ask about what kind of change of life might offer us a bridge to the future is Figure 5. It represents a smooth continuity of organizational processes, not mathematical shapes, and is called *nature's integral*, the classic shape of how nature adds things up. There are similar mathematical shapes, but they do not say much about what is happening to produce them. Asking questions about what produced it is the object here. The depicted organizational growth process is initiated by some *germ* of system design,

Not immediately obvious is that needing *Power for Takeoff* and then the longer period of *Maturation & Fitting-in* shifts the power source from one process to the other. The power first made available by *Innovation* is then used for *Coordination* culminating in readiness for *Engagement* in the *New Life* ahead. So here we have another case of seeing what happens and finding it hard to quite understand how it happens. However, we still urgently need to know how to save our world from our long-held blindness to the limits of our inventions! So we need to find how to be practical, find a new way to innovate, and tap into any kinds of work not yet widely recognized, offering plausible strategies for how to do it. Perhaps most importantly, we need to study all related natural or human designs to find cases that might show us better ways to do it.

Perhaps the main pattern to focus on, the one getting us in trouble, is why it is that some growth systems smoothly turn from growing their power to exploit their worlds (growth stage - A) to instead harmonizing their systems with their worlds (growth stage - B). That "change of purpose" seems to sometimes come from a system having more "growth-pains" at the same time it recognizes a new kind of "growth-opening," from making things big to making things work. As systems grow, they have more maintenance needs; things get more complicated, and new relationship needs are further away.

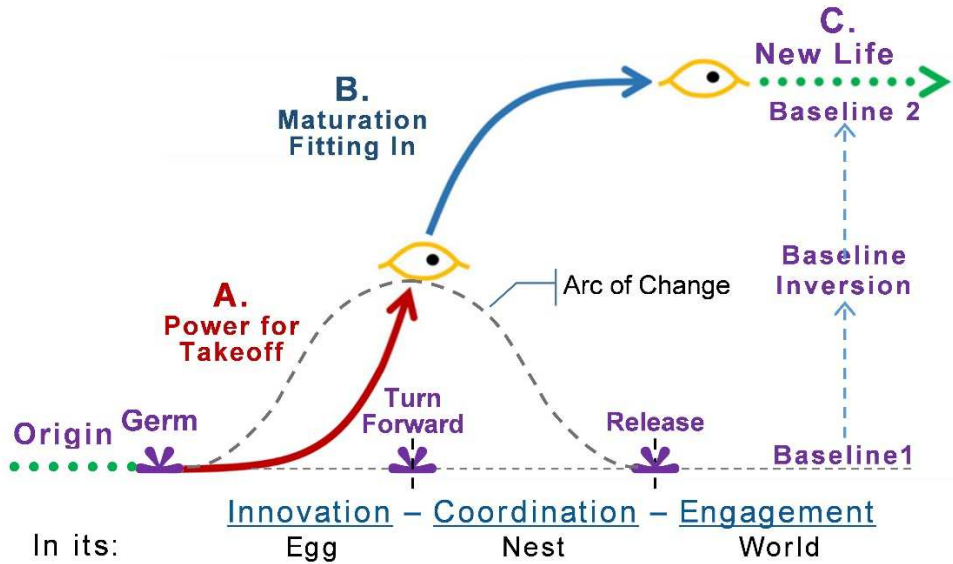

Figure 5. Nature's Integral – Pointers to developing long lives. Each main period of development, A, B & C, would be made of various stages building on the ones before that produce growth and transformation, just not shown here for generality. They are what to look for as the actual pathways of system change in any particular case. One sees these successive stages in work on any project, new relationship, or education, combining to form the integral whole.

These contextual triggers might guide a system's animation to "turn forward" toward perfecting what growth built. In effect, that would naturally divert resources formerly used to grow the system to caring for its needs instead. People do that at various stages of life, such as when we come to an end of growth, ready for adult life (moving on to engagement - C). One of those needs for any system that responds to its internally felt needs to stop growing to invest in maturing to survive is to have resources to continue to be creative, take on new roles, and change with its world.

The learning challenges to achieving a smooth transition to a better and more sustainable way of life, mentioned in §1.2, are large. They will surely not be all those confronted. Crises inspire innovation of new kinds, though. In this case, learning how to do it is even in the very direct interest of the major economic sectors, led by the best educated and successful people on Earth. That they seem blind to the threat to themselves and everyone else, of continuing to maximize the system's growth till it breaks down, is another of the fascinating ironic puzzles. That blindness seems to be the specific reason the system is not responding to the increasing needs of the system to care for itself and its world.

Part of what we will most need to steer business choices is a way to calculate profits and losses holistically, not selectively as at present. There are ways to begin doing that, estimating the full spectrum of ecological and societal losses if overwhelming the Earth (Henshaw, 2011, 2020) and then qualify responses and distribute the optimal level of funds. That could distribute costs and benefits globally and in everyone's interest. The math to do that kind of distribution seems feasible, at least. But, for the moment, the extraordinary pushback from nature to our relentless expansion and interference with nearly all of life is a clear signal of harsh limits ahead if we do not collectively act.

### The three beginnings of transformation

alpha, beta, gamma – α, β, γ

The three system development periods – A, B, & C: *embryonic* and *maturing* growth followed by mature *engagement* – are often very noticeable and consequently easier to study. The deeper event of transformation that begin the three longer periods are here being called *germ*, *turn* and *release,* and in Figure 6 below, given symbols alpha (α), beta {β}, and gamma (γ).

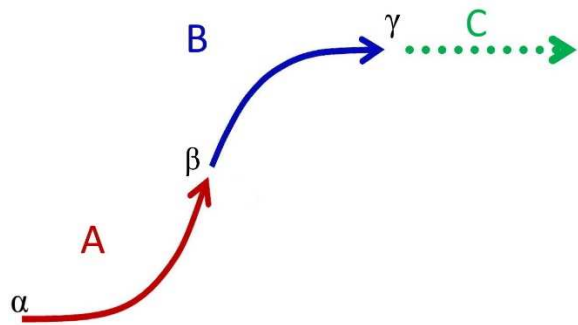

Figure 6.  A, B, C – α, β, γ    The labeling of syntropic development stages.

These transformation events initiate the main changes in the directions of development, sometimes very noticeable as events or too fast or small to see or record, marked only by differing whole system behavior before and after. For example, the spark of interest when a future couple first meets may either be very noticeable or easily dismissed by either or both people; those connections set the wheels in motion for a building process in which their experiences will affirm in context or not.

Observable or not, what is special about them is they appear to involve deeper complexities on multiple scales of organization, "differences that make a difference." While those transformational changes in direction are generally faster, simpler, and more transformative moments of change, they may well not be simple at their scale. At their scales, they may have development shapes like Figures 5 & 6, consisting of processes with as much complexity and variation.

The puzzles of *alpha* shifts come from asking how they work, like what allows a *single sperm* to fertilize an egg? It seems to require a mysterious smaller molecular scale beginning, middle, and end organizational processes. The hint that an organizational rather than statistical process is involved is that somehow only one sperm seems able to do it, so it's not an open door. Perhaps the egg responds to only one sperm, somehow *selected* for allowing in. Life is unquestionably an amazing engineer, making it plausible. It's s rule that may also result from selective *attrition* due to eggs that allow more than one sperm to enter failing to successfully develop, or both?

A similarly curious *beta* shift is the variety of ways people decide whether to mature or let go of new personal relationships. They all start with an experience of growing out the first pleasing connections; however, as all growth does, that first phase is outgrown. That happens in many different ways and then follows many different paths. The ones that come to matter most to us seem discussed and experienced as making the *turn* to deep, lasting relationships at a particular moment, somewhat by surprise, erasing all doubts like a miracle. It changes to protecting and maturing the new roles rather than just riding high on them. But how in the world does something suddenly do all that at once? External approval of family and friends often matters; sometimes, both mutually reinforce the deepening of shared life experience, attachment, and feelings! These life paths vary widely, such as being passed up and rediscovered long after.

The variety of *gamma* shifts is at least as varied, also seen in how people start their adult lives in so many ways, take so many different kinds of lives, then change and shift between them. It makes the description of maturity as *homeostasis*, meaning ever-returning to center, seem to be a great oversimplification. These critical organizational changes may be tiny and instantaneous to trigger the whole system shifts in the development direction they cause. They may also be glacially slow changes in larger scale balances triggering deep organizational change somehow. They may even be simultaneous, with the whole system momentarily coordinating large and small scales; who knows? It is a little like speculating on other universes.

## 2.1 Reading the Signals

> *"The Lyf so Short, The Craft so Long To Lerne"*
>
> *– Chaucer*

Here we explore what growth system steering (self-control) is in practice by learning more about reading the already fairly familiar signs that nature posts along the way. Another version of the same simple S curve diagram, to use as a map of what to look for, shows why we don't see what is happening at first (Figure 7), will prompt leading questions, and help identify more non-verbal cues. Very small signs can signal big things, and very big things can imply little or nothing. Learning to see these meanings is an ancient language for reading the sources and behaviors of emerging change. Unfortunately, our modern world's focus on finding rules of control rather than on how to notice how natural systems work by themselves causes that knowledge not to develop.

We all read signals of things changing we might need to respond to all the time, sometimes it is easy and clear, sometimes not clear what it means, and sometimes we are much too slow in seeing or responding. So to learn more about it, we could look more closely at where the signals come from and share our experiences of learning what is more important. The best place to start a discussion is to ask a group an opening question. That might be: "What have you noticed?" or the more detailed"Tell us about your fruitful, satisfying, disappointing, or funny experiences with reading signals of change and how you responded." Noticing *changes of life* that need to be left alone are at least as important as changes that need support or might be a threat.

We might read a signal of a change of heart and instantly forgiving a person for a great wrong, which turned out to be a mistake. I also recall asking a banker about managing an ever faster, changing, and more complex world, who said, "Oh, W E can handle it!." It clearly suggested that bankers had no idea what was coming, a surprise only confirmed by the wider systemic failure to respond. I also continually learn and relearn social and physical skills, watching the flow engagements for signals and ways to respond.

As we search for signs and cues for response while watching things happen, we notice what gets our attention and how to better time and measure our responses. We also notice the opposite, seeing what are only distractions and need no response, and the opposite, what we very often miss. As with many kinds of observation, a higher level of perception comes from being able to later recall the contexts of events not initially noticed.

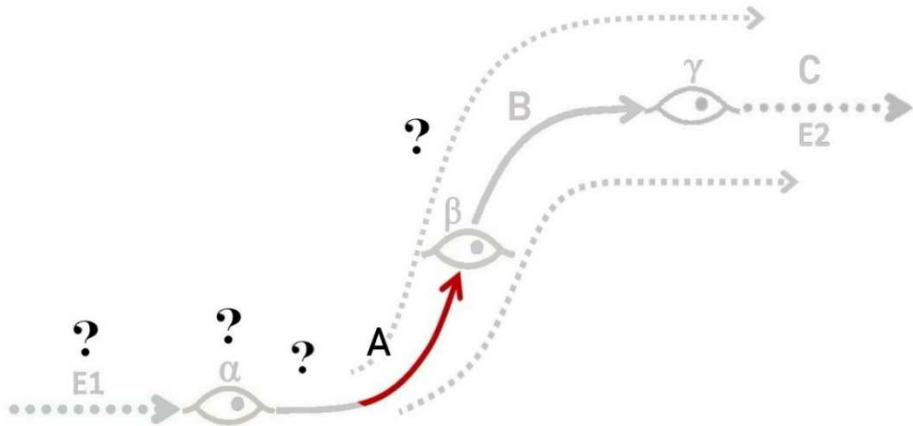

Figure 7. Questions raised by our narrow view of emerging systems: We are likely to only first notice the intrusive growth period represented by the red line and not notice it as nature being very "pregnant," with ALL the associated interests, questions, and concerns about the internal animation and expanding enabling context that comes with that. – α then A, β then B, γ then C -

It does seem odd to think of remembering later what you didn't notice before, but it only takes remembering more of the background than first noticed. It makes retaining *raw impressions* and *holistic feelings* that one can reconstruct important for understanding the meaning of events. It is still the kind of recall that needs checking, like a good hypothesis. The verifiable connections that kind of recall helps one trace make both validation and falsification easier, and it also raises many often unbiased new questions.

If one can recall contextual impressions of events, sounds, and smells, we can think about the prior context (E1), the germinal event (A), the startup period (the grey part of 1), and the later stages of what may have or might happen (B, 2, C, & E2, end context). Those would all be part of the natural provenance of the causal connections. To think about it holistically, first, think of all those connecting parts as an eventful flowing progression, like a story, and adjust the *arc*s as one would the shape of a story as you learn more. Then look at different parts, reviewing what you can recall, find out, feel, or imagine what has or might be still happening and the effect of or on its context. You would expect that odd interruptions or departures from the general flow could be important or not. That is, of course, because life cycles and information both tend to be eventful in surprising but different and the general symbolic model is only for helping you look into the reality.

Many people know about a heightened perception of the details of things, from life-threatening or world-changing experiences, suddenly seeming to see in slow motion. That comes from the mind shifting to a higher speed of recording our natural senses of what is happening. Artists and performers may have more of a natural talent for it needed in their work. Still, anyone seems able to learn to watch it as it happens occasionally and then extend that ability.

Slow motion perception is particularly valuable in quickly noticing things, such as the earliest signs of beginnings and endings. What catches one's attention is often seeing something surprisingly out of place. It might be an unusual calm in a usually noisy and busy place or signs of unexpected changes emerging, as suggested in Figure 7. As we notice new things emerging, the information always seems late because it signals something having an unseen history and likely expanding future. So to catch up with the past, it would be nice to know how to quickly replay the recent past in slow motion, retrieved from your recorded contextual awareness. That would help identify any urgent information about where the emerging change came from and is going.

That frequent perception of emergent change is an autonomic semi-cognitive response to the dynamic features of emergence, recognized as a non-linear progression, even if still small. It could be a pencil rolling off the table or a puzzling look from someone as a sign of rapid change. That natural ability to quickly become alert to change is something we can improve on, too, by watching for the signs and being more ready for the interruption. A good example is working in the kitchen and ready to jump if a knife falls off the counter or while driving, ready for a person to suddenly appear where not expected. One wants to instantly act but also have a speeded-up presence of mind to not panic.

We also notice emerging action and inaction as cues to act or leave things alone. It is another reason to have a speeded-up presence of mind when noticing significant changes (or lack thereof). The most meaningful things to notice may be about things to be left alone and allowed to develop naturally, without interference. Similarly, it is often good to mull it over for a while when one has a bright idea. Letting it jell and slowly sort out connections in the context to make or avoid before giving it structure and purpose.

**The midpoint of growth**

The classic example of a pivotal change to focus attention on is the midpoint in the S curve of growth, its "inflection" or "turn forward" point, 'β.' The curve reverses curvature, going from the **A** period of *takeoff* to the **B** period of *landing*. Often missed is the profound internal system change, as its driving purpose from multiplying its power to adapting to and exploring its new world. It can be a simple or a quite dramatic event depending on the change of environment that goes with it, <u>and not seen at all in the growth curve</u>. The new chick and new child run out of space to then emerge from their egg or womb as the protected places for their periods of boundless explosive growth, to find a bigger world to adapt to and make their own.

For the farmyard chic or new child, the transition is quite dramatic and cues rapid responses from their environments, as if left in the lurch, nearly exhausted and cut off from their food sources and unable to fend for themselves. That change is also a dramatic shift from focusing on *internal* to focusing on *external* relationships. It not only occurs at the limits of physical growth for new systems but also for the growth of personal relationships when they shift from focusing on themselves to focusing on relations with family and friends. There is a shift from focusing on oneself to finding one's center in groups of friends, work and businesses, or community and organizations.

The important takeaway is that this change in new lives, wherever they came from, is to make their own turn to where they are going. These are critical formative stages for the new life of any emerging system, a business, society, individual, or other emerging life. So for people and their organizations, it is critical to be able to observe, for example, what is growing right and wrong around them, and have some idea of what to do. These pivotal changes in direction have large, lasting effects, and every new system or life is steered from the inside and can only have learned how on its own, nature's test.

### Noticing S Curve Transformations?

Did you take notes on the main events that occur at different points on the S curves of emerging new lives and transformations? Did you think about what they mean for what is developing and the world it is developing in? And, Why are they important to notice? Below is a simplified version to refer to as you read the notes on what things to notice and talk about with others.

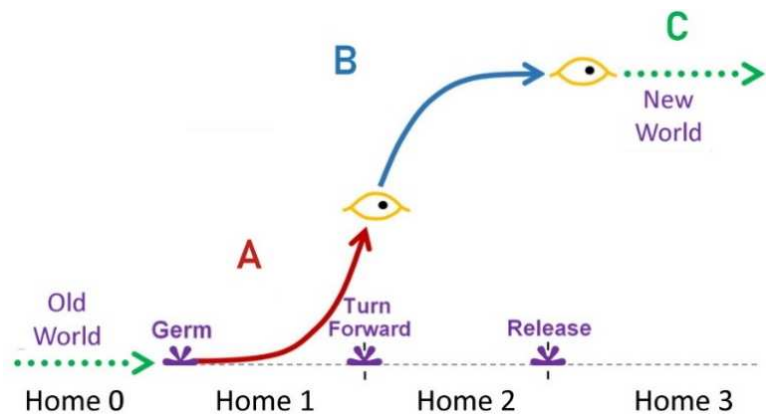

Figure 8.    Small S curve figure Key

Before and After ? – The Context

1. What were things like **Before** ?           Fully describe one example?

2. What will they be like After ?              Think of several you could describe ?

Hints:  Do what is important before and after form different patterns?

Holistic Natural Systems - Design & Steering

When Growth Starts ? – First Hints

3.  How do we notice what is brand new ?     What are several examples ?

4.  Do they change their world ?     How can you tell ?

---

Hints:  Things out of place ?  Things happening again ?  New faces ?  New questions ?

From beginning toward ending ? – Mid-Point Turn

5.  Buildup coming to an end?     What are examples ?

6.  and perfecting things starting     What are examples ?

7.  Does that change their world ?     How can you tell ?

---

Hints:  New patterns of change ?  Looking to the future ?  Changing environments?

Coming to an End and Moving On? – New Roles and World

8.  Approaching perfection?     What are several examples ?

9.  Getting ready for new places     What are examples ?

10. How does it change its **old world** ?     How can you tell ?

11. How does it change its **new world** ?     How can you tell ?

---

Hints:  New patterns of change ?  Looking to the future ?  Changing environments?

**Parts moving altogether.**

One of the important properties of systems working as wholes is that measures of their parts tend to move together, Figures 9 & 10. For example, figure 9 shows major indicators of the world economy: GDP, consumption, pollution emissions and accumulation, energy use, and GDP energy efficiency. Remarkably they are all moving in closely constant proportion to one another!  That exhibits their *proportional coupling*, a clear indicator of the world economy's organization, causing it to work as a whole. At first, one might think that cannot be, as the research papers all show that the various parts behave very differently. But, that is what globalization overcomes. Market forces steer every part to find the most useful way for it to fit, driven to maximize returns.  So, rivers of innovators, workers, businesses, and investors vie to provide the services the world wants and get the highest price for them to have growing investment resources. So the national GDP curves jump all over until you put them together.

How the curves move together also shows that the system's organization is unusually stable, with measures of the whole having near-constant growth rates (i.e., rates of doubling) that are in constant proportion to each other. The close fit between the data and average growth-constant curves indicates that it is real. It shows the economy to have fixed rather than responsive steering, too, a fatal flaw in a world of rapidly changing conditions and threats, almost sure to be steering into ever worse trouble and inescapable traps, like having no achievable goal but ever faster change.

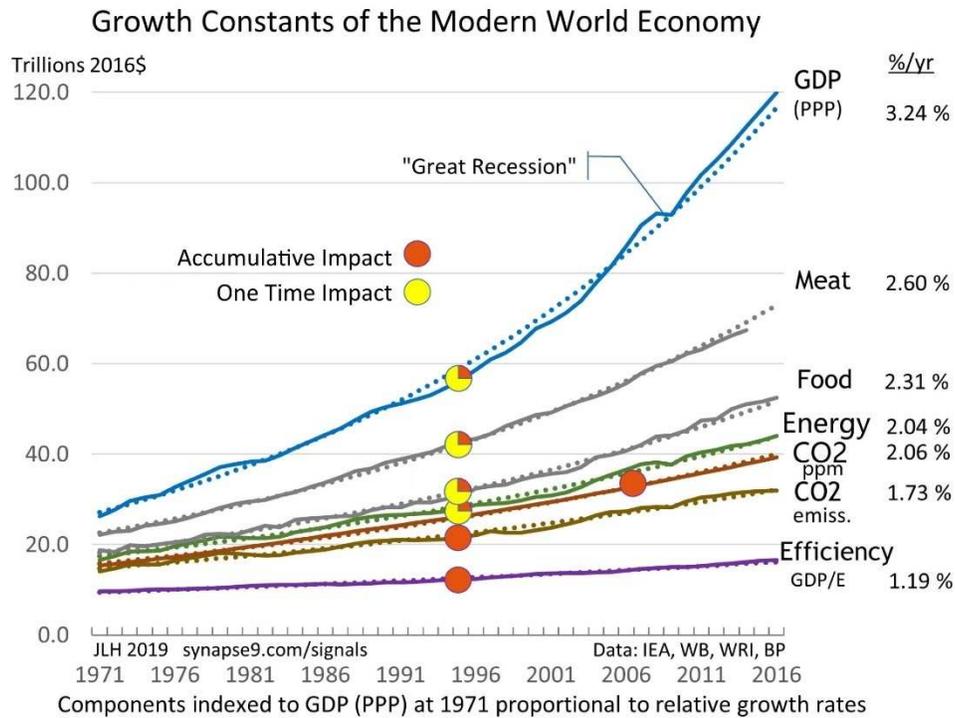

Figure 9. Global economic steering curve, showing the close to constant proportional coupling of exponential resource and consumption impact curves.

That fixed steering, of course, is for maximizing the growth rate of GDP, investment, profits, consumption, and failure to slow the steady exponential growth rates of lasting impacts. Nearly the whole world focuses intensely on climate change, but we appear not to have yet considered changing the economy's steering. As a result, the latest atmospheric $CO_2$ data (the direct proportional cause of the rate of warming) still shows it rising at its highest ever exponential rate (Henshaw, 2019).[21] So what would it mean to steer the economy in some other direction?

Since investment builds the future directions of the economy, i.e., steering how the economy will turn and where it will go, new directions for investment will change where it is going. That would involve having more than a single variable objective. For example, it would make sense to steer the economy for some safe harbor to protect it as it transforms. The current plan is to keep it multiplying the impacts that threaten itself and the earth. To do that, investors need to a) understand the difference, b) develop a plan, and c) have the plan give them the social, cultural, career, political, and financial motivation to act in the common interest.

That not only seems necessary; it also seems possible. The people with high educational, professional, and social status are in charge of steering the economy on its fatal course today. If they saw the chance to correct the error of steering the economy to maximize the growth of its already overwhelming impacts, they would respond out of wounded pride, if nothing else. That would then motivate the technical teams to expand sustainability to investments using reliable measures of accumulative costs and benefits (Henshaw, King & Zarnikau, 2011; Henshaw, 2014). Something needs to relieve the system's ever-growing distress, and it would be better to be doing something right than wrong.

Figure 10 shows word use frequencies for some terms of distress in English books by Google.[22] Selected are common terms for personal distress that have been rising together as probable evidence of systemically increasing societal distress since the mid-1960s. Their roughly parallel compound growth period is from about 1975 to 2011. The fluctuation after that suggests something hit a limit, but feelings of distress remained high and unstable. There is enough detail in the shapes to make them look possibly associated with specific cultural changes. That the curves and their fluctuations move together, suggesting the curves reflect whole system behavior, might make it hard to find what is involved. The pattern seems extraordinarily regular and systematic. Interpreting a long period of exponential indicators moving together, indicating they reflect a whole system behavior. That the pattern begins in the

---

[21] Scripps Global average Atmospheric CO2 ppm data, updated in May 2022.
http://scrippsco2.ucsd.edu/data/atmospheric_co2/icecore_merged_products

[22] Google Ngram of societal distress terms moving together- see curve back to 1920 to more stable period
https://books.google.com/ngrams/graph?content=pain,anger,anxiety,hate,hurry&year_start=1920&year_end=2019&corpus=26&smoothing=3

mid-70s is quite significant, suggesting some lasting change started then. It could be lots of things but interestingly was not in the mid-50s or 60s when there was also emerging awareness of dangerous crossings of environmental thresholds. From personal experience, the long rise of modern anxiety did seem to start in the 50s and 60s. Evidence suggests that something more contagious or disastrous may have emerged in the 70s. Short lists of recalled sources of general distress for each period are in Tables 3 and 4.

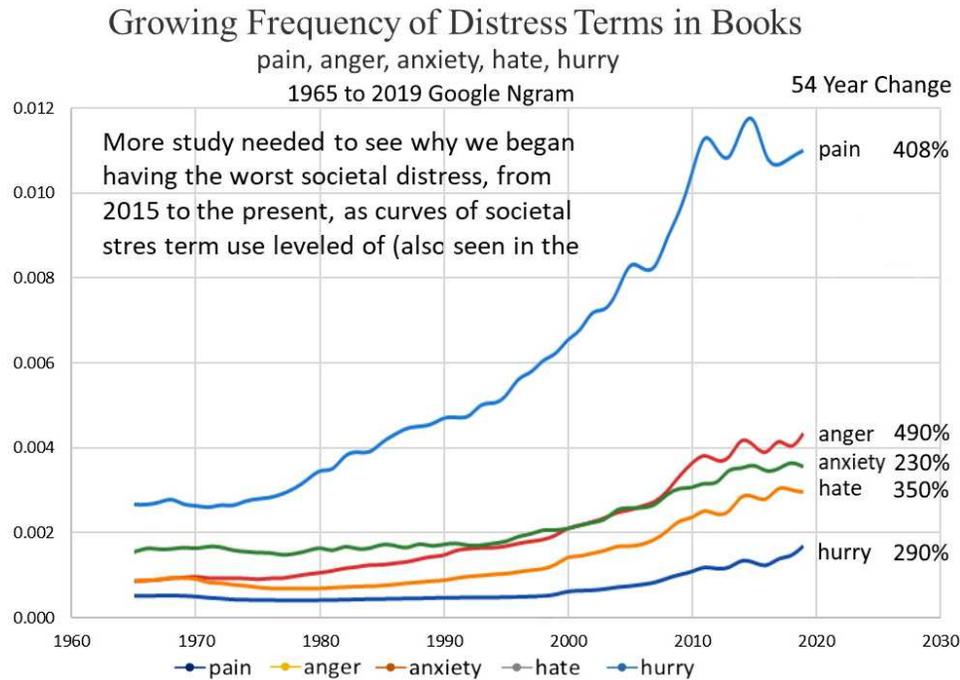

Figure 10. <u>In books, a 350% rise in terms of distress – moving together:</u> Word frequency in English library books scanned by Goggle for the distress terms: pain, anger, anxiety, hate, and hurry, shows steady growth then fluctuation.

- Nuclear bomb threats and Silent spring
- Kenedy assassination and Viet Nam war
- Race riots, social revolution, & anti-communism
- Shareholder value took over the stock market??[23]

Table 3.     Emerging societal distress in the 50s & 60s

---

[23] Did it start with "stakeholder value?" - Graph at: https://synapse9.com/issues/GDP-WageHistSMb-fig.jpg
- The 2016 research notes: https://synapse9.com/signals/was-shareholder-value-what-did-it/

- Severe recession and inflation
- The energy crisis
- Beginnings of political hate movements.
- Big rises in urban crime and violence

Table 4. Emerging societal distress in the 70s

Other features of the Figure 10 curves are a semi-regular fluctuation, long regular escalation on four of the five, and an abrupt shift to matching large fluctuation in about 2011. Obama's first election was in 2008 and second in 2012, followed respectively by the biggest rise in societal distress and the first big drop. So that does not seem to make sense, as Obama was awarded the Nobel prize for restoring a sense of world peace. At this point, when running out of guesses, it is time to question assumptions, look around the context for anything neglected, and find some authoritative studies. The evidence so far is that something has been ratcheting up the levels of distress throughout the English book writing world, about as regularly as the ratcheting up of the economy. That the pattern had a specific beginning, coincident with the stock market change from reflecting business value to shareholder value, makes it seem plausible that is one of the drivers.

A hint also comes from the sharpest rise in distress coming in Obama's first term – when he was celebrated for fostering world peace. Ironies like this are often excellent evidence of looking in the wrong place. Perhaps those feeling most distressed by the rapid deindustrialization and re-socialization of America happening at this time and also by the election of the first Black president are the people feeling the pain. That might not be who the readers of this work would think of first, as it is not an intellectual community. So it is a wild guess, but this data may capture America's vast writhing political polarization and perhaps in the UK and elsewhere. The longest and most rapid accelerating polarization of the US economy[22] would be guaranteed to disenfranchise many real people. It might disenfranchise whole sectors of society, which would certainly deserve to feel its pain.

That is not what was first assumed at all. It does, however, fit both 1) a social blindspot and 2) a hypothesis based on *reading stories into the non-linear continuities of the data,* prompting the recall of additional relevant data. Is it plausible that the ever more distressed English book writing community located by the Google data is the political far right? They most vociferously voice their pain and have done so for quite a long time. Still, it is a surprise. We have learned a lot from the study and come up with unclear indications to help keep us looking for more and from jumping to conclusions! We would assert that this very constructive conclusion comes from the method.

**Spiraling global forces**

What is happening is always the question. For example, say you are a senior director of the world climate response team, and you see the following graph of atmospheric CO2 (Figure 11). Say you have seen the graph repeatedly, year by year as the dots on the curve proceed in a direction quite different from the plan. What are your choices? What is top on your list? Might you ask, "What plan are they following?"

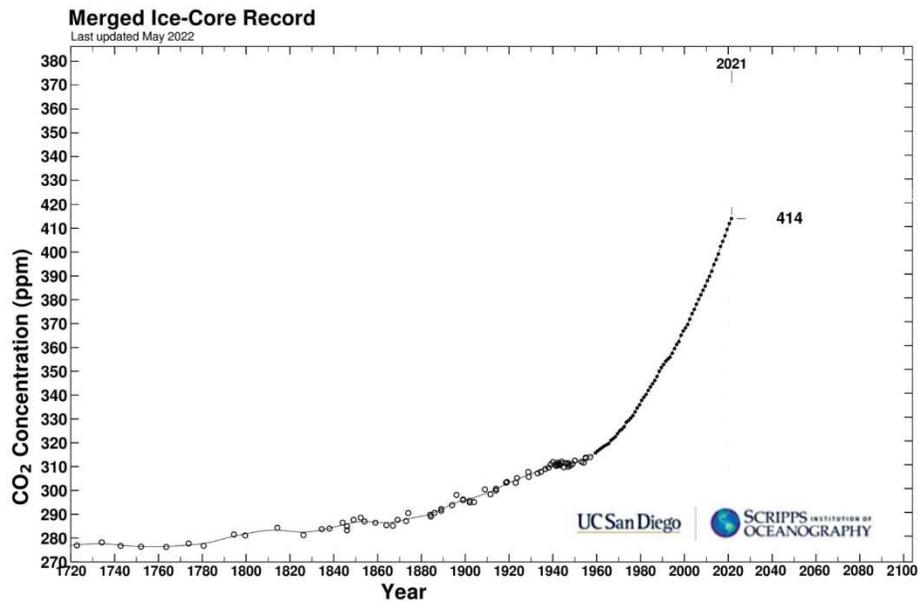

Figure 11. The Scripps 300 yr global atmospheric CO2 raw data: How do you read it? Does this help indicate the true root cause of our problem? Where does it look like the next dot will be?

## 2.2 Transformations

How brilliant business innovations emerge is often from following a passion, not knowing where it goes. Modern science came from Alchemy, in something of that kind of way. The transformation of our out-of-control world economy may well get ahead of us and so involve a lot of catchup. So, now seems a good time to study the available options and not fail to ask, "What If We Get This Right?"[24]  Nature *does* often guide emerging systems to get it right. The main message of this paper is that there are ways to study that, learn how to read the signals and make the right responses. Circumstances will differ, of course, but studying the basics, considering their neglec, is perhaps the most important place to start. That is, learning to be observant, then using it to see how the systems you know do and don't work.

---

[24] Dr. Ayana Elizabeth Johnson - https://www.ayanaelizabeth.com/

Finding words to use is part of that. Watching the flows and making them into stories helps. Mining the inherited meanings of the common language to tell us about the things we notice and wonder about is perhaps the best way to connect our language with its deep roots. That might also get the closest to finding and responding to the root purposes and visions of the coupled human-natural system as a whole.

People naturally learn nature's creative processes and form the foundations of their understandings in life, immersed in the environments they come to know. These include people immersed in everyday work, community, or organization relationships. More generally, every community and local ecosystem is also a systemic *hive* having *low degrees of separation,* so the responses of most parts are to the whole, making the whole tend to be cohesive. But, of course, the big problem we face of blindness to our impacts, using rules for remote control of systems we little understand, also calls for a deep moral and historical search. It would be good to know why and where we began tolerating the self-deceptions.

Advanced learning about natural systems seems to take a different observation method, imprinting the mind with immediate impressions of patterns, relationships, and change, expanding on that natural ability to observe and directly study our realities. We can also improve that ability, too. We do that by learning to see the fine nuances of differences in any subject as we do with the hour-to-hour changes in our children sometimes. Almost any subject of fascination will do. That seems to be what Goethe's observation technique did for him (Goethe, 1996). One can even do it with one's feelings and reasoning about them, using each to freely and openly respond to the other, perhaps discussing discoveries with others. Many good observers, like scientists, entrepreneurs, and artists, rely on honest and self-critical perception. Unfortunately, it is still hard to fend off some of our cultures' biases, myths, crazy politics, and obsessions. That our economic culture focused its attention on how to control each other's minds and the natural world set back our ability to appreciate how nature works by itself.

At the atomic level, we cannot observe what happens nor see if atomic forms have internal organization. There seems to be no useful information yet, but if energy conservation still applies at that scale, some behaviors may not be statistical but developmental. What is most important to people is not the theory but that these patterns apply to our lives and activities, all our large and small activities beginning and ending with accumulative organizational steps coupled with an environment. We will always struggle with the resulting perspectives of systems as seen from inside and out, always being quite different. Given that a system's internal organization is a self-referential whole, it will generally not be visible from the outside.

We live in a world where perhaps the most important causes and effects, those that develop from whole emerging systems and animate change in our world, are generally found to have hidden interiors and be out of our control. So, to appropriately respond to those emergent causes or the struggles systems that need help, we need experience reading the cues and signs of what they are becoming or how they are struggling. We become somewhat successful in noticing the smallest signs of change, watching our children, friends, partners, etc. It is also the core talent for having "social skills." It seems to be what indigenous cultures teach and how animals get along in a complex world, born to be curious about the hints of change in the world around them. Earlier research (Henshaw, 1985, 1995, 2000, 2008) pointed to a subtle sign of important change that could be seen in time-series data or noticed by a sensitive listener or observer. That is reading the non-linear dynamics in the background as changing derivative rates of change. Sometimes it prompts one to notice things hardly noticeable before.

Modern cultures teach awareness of that non-verbal language of change for personal matters, like home cooking or personal relationships. Still, in developed societies, education long focused on teaching rules, names, and concepts, though that is changing. Learning about whole emerging systems from abstractions still seems to be the general rule in business, finance, and most sciences, where a precise "bottom line" is important. The problem is that it teaches about systems detached from their natural contexts, starving the learner of questions and intuitions about what is happening in nature. It is another reason people might feel uncertain about natural change and feel a need for control, not taught to be observant.

## Breakthrough Transformation

There is a major risk in even posing a search for "breakthrough strategies" of simply falling into our dominant world habit of solving problems by finding things to control. That very organized way of life has given people considerable power, influence, and control over each other and nature. Unfortunately, it has left us as a species totally out of control. We created societies organized to rigidly continue to multiply their control.

That system of life is destroying life, an entire planet, and causing us to avoid responding to the natural urges to use our wealth more wisely. So the first urge is to look for how to use wealth to heal the wounds we feel and see all around us. It is indeed possible and not altogether wrong, but it is also a step back into using our power to exert control again. So the question also needs to include how to get out of those runarounds?

So to start over from basics, nothing appears meaningful except in context. *Runarounds* seem to be questions about systems caught in loops of *cause-and-effect*. They might well be headed somewhere on a spiral path or not at all.

From the widest view, we have built an amazing new kind of life on earth, do not want it to fail, and need a healthy world that is much more naturally self-controlled. What that looks like is *the start of a plan.* Our current major threats are self-inflicted, seemingly caused by our fixation on using controls to concentrate power and profits, and for the longest time not noticing it was and would destroy the earth!

We developed those habits over the last 60,000 years since widely shared conceptual thinking began, and around 12,000 years since that powerful way of thinking created urban centers. Then 2500 years ago, that simplified way of thinking about power to gain more power turned into technology and science. After a dark age, it reemerged in the Renaissance, then 240 years ago helped us begin multiplying $CO_2$ and then find more and more efficient ways to do it (Figure 11). Every growth system starts somewhat that way, of course, but can we learn nature's trick for taking care of things on a global scale instead of taking power over them, the fascination we let reemerge in a bigger and bigger way after every failure of it before?

It seems to always be inspiration that somehow takes over whole systems that produce transformations, though. So, we do need to not shy away from having inspiration. But, this time it would need to help free us from an exceptionally old runaround, so we can continue our original journey, keeping the skills for creating and organizing things that work.

Like our bodies and minds, all systems need fixed and moveable parts so that they can go places, is one way to say it. So system designs need tension and release to maintain balance, stimulus & response to remain creative and distributed, and so the wholes build on foundations and evolve. Those are structures of system self-governance, relying on various forms of internal balance, sometimes called *polycentric governance* or *tensegrity* (Ostrum, 2010; Turnbull, 2022).

The following citations only scratch the surface but offer some insight into the imminent risks of pushing systems to points of fragility, collapse or internal decay (Bell, 1971; Tainter, 1988; Chew, 2007). Many times before, cultural decay has led to dark ages, seemingly driven to unsustainable complexity and collapse, as Tainter observed. So it should be a serious concern. The complexity of systems generally cannot go backward and becomes unproductive too. So it's quite problematic that modern economies have been relying on it for some time.

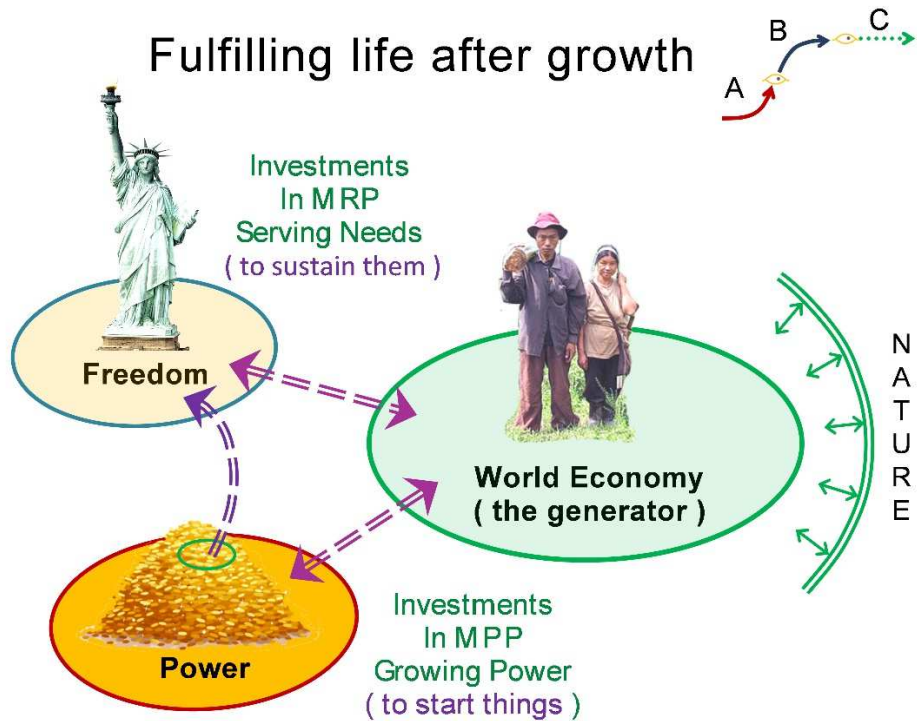

Figure 12.Systems that outlive their growth: start with a maximum power principle (MPP) to rapidly build their forms and places, then a maximum resilience principle (MRP) to care for what they built, their lives, and world

So, hopefully, these sobering truths about the kind of threat we face, and the opportunity to "start a new life" offered by learning from familiar examples of how to perfect complex system designs that develop by growth, will both help bring about the inspiration and clear enough minds to let us creatively shape our best response to our impasse (Figure 12).

Figure 12 above is another simple concept intended to help think through the creative steps to take for releasing it from the trap it is presently in and transform the economy in nature's favorite way to creatively prosper for a long time. The economy could continue to work smoothly as the generator of wealth. Well-informed people with vision could also help the community of well-educated, successful, and communicative people who manage the world's businesses, governments, and finance to understand their new job. That is to take care of the natural world we are part of and the homes and societies we built, rather than continuing to manage our multiply exploitation of them.

## 2.3 Healthy systems

There is a useful shortcut for understanding natural systems once one has a grip on the basics. Of course, one will always need to go back to the basics again and again. That shortcut is to read systems and the events and changes of their life cycles as the great stories they are (Henshaw, 2018). One can read the progression of the flows as *the arcs of their stories* about the flowing developments in their experiences and relationships, paying close attention to smooth *takeoffs* and *landings* and what inspires them. Those are *nonlinear* features of emerging natural design that are hard to fake and important to explore. It also helps make the intuitive guesswork that goes into the stories one reads into them *reasonable hypotheses to check out*. How much one does of that or not, the storylines remain as markers of where you might want to check. To bring out the natural systems that anchor their meanings in their contexts, use terms and language that help direct attention to the natural processes, relationships, situations, and experiences that our languages developed to convey. Of course, that includes shared social and intellectual experiences, too, as they are part of the natural systems world as much as physical experiences. Part of telling real stories is keeping them real by noticing and weeding out confusing terms of social and racial prejudice, misinformation, false authority, etc. Those do sometimes creep into our language if they are circulating around us.

### Designs of homes

How families make decisions in the collective interests of every member and the whole appears to have been the major change from human tribal to home + community cultures, becoming foundations of human culture and societies seems to have developed toward the end of the Stone in the Bronze age. It seems exemplified by the design of the proto-Greek Aegean Hestian home (Figure 13) and its apparent worship of home life central to their culture (Henshaw, 2022) from which our homelife cultures descended (Dinsmoor & Anderson, 1973).

Much the same strong allegiance to the unity of the whole if seen in groups of good friends. It also sometimes characterizes mature organizations and businesses that, as they perfect their designs and mature, turn to using their profits for engaging with their worlds. That is the ideal natural S curve culmination of long-lived natural systems that, as an ideal design, seems to have allowed the emergence of complex life and reproduction long ago.

Life is hazardous in any case; things happen to prematurely break relationships, and the lives of individual systems never last forever, though large collective systems often continue to evolve and long outlast their parts. Human communities exhibit some of the "all as one" features of homes and close friends. At larger scales, more organizational diversity and complexity develop. More people follow interests separate from others, with the community still working as a whole as the parts fit together, having a common culture of connecting differences (Figure 14). That is quite visible in most small and large communities and even big cities like New York. On those scales, the common organic cultures also host compatible and antithetical subcultures, cooperating by staying within bounds of tolerance, except when they do not, but rarely ever breaking up. After years of global travel, neighborhoods are more like a *salad bowl* of different cultures.

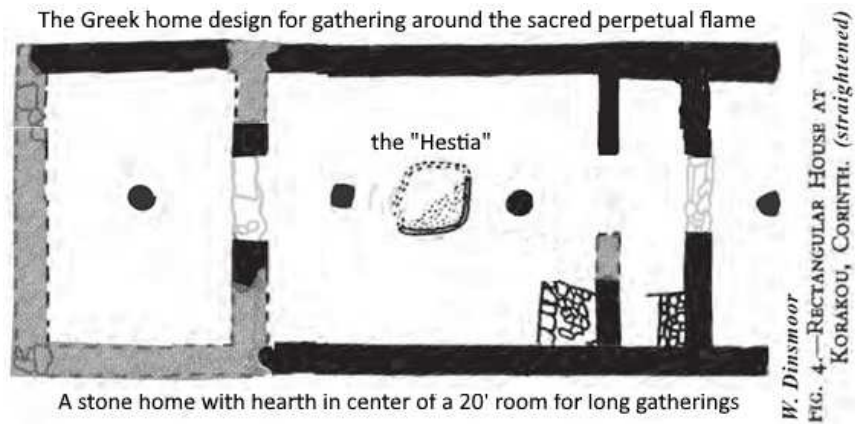

Figure 13. Natural system centers have physical or organizational enclosures with openings that provide easy access to filtered inputs and outputs, serving as homes for their hives of internal relationships that organize and energize their lives.

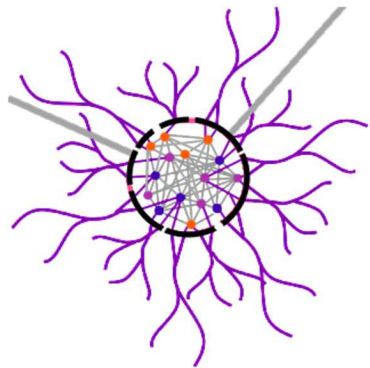

Figure 14. The organization closely connected human centers, Homes, Gatherings, Neighborhoods, Communities, and Cities: the Hives of culture that become Nodes of connection in larger networks

The interiors of homes, businesses, communities, organizations, and economies all contain and energize thriving *hives* of relationships. The term *hive* refers to the connectivity of a domain, usually the interior of an organized system, but used to refer to the small worlds of close association that serve as *nodes* in *networks*. So within a hive, all members would have direct connections and zero degrees of separation. So, for example, the hives within a network node might consist of the people in a family home and from their close association collectively responsible for messages sent and received on the network. It is just one of the interesting dynamics connected with *centers*.

Centers are the basic unit organization in the design of life, as *centers of control – <u>with</u> – open connections*. In architectural design, they are visible at every busy street corner or local hangout, spontaneously forming and dissolving as the "attractors" that puzzle complexity theory we all naturally know very much about. Of course, there are hives of connection of many other kinds, such as those with and without networks connecting them and composing the complex textures of environments. There is much more to be said about the role of centers in our lives, how we build and care for them, and how we experience the freedom, security, and connections they give us.

## 3 Discussion

> <u>Reflection</u> – *Will people will have the inspiration to survive?*
>
> This study has led to the conclusion that for whole complex systems to change as wholes and set out upon a transformation journey following an S curve to a change state, they would need to be inspired as a whole by an innovation that extends their origin story, soul vision, and original purpose, taking it forward.

The views on what will become of the present confusing state of the civilized world, to both those who closely watch it and not, varies widely, despite the broad agreement we are in a very threatening whole system crisis. What is necessary is going to happen, whatever that is. Whether people can or will do with it what is possible is considerably more in doubt. In a rising crisis, reaching a general state of heightened suspense and nervous anticipation before doing the right or the wrong thing seems both the most and least promising sign of resolution.

What Putin did is one example, feeling he had to act against the forces of history, working up the suspense and his courage to violate every principle of goodwill, then launching a major military incursion bent on erasing the homeland of Slavic culture he seemed to think had betrayed him. How Trump lost his sense of reality seems to exemplify the madness that comes from opposing inevitable forces as well, becoming desperate and spreading false stories to attract others, also fearing for the future. We should not follow those examples. Maybe we are lucky to have them to warn us.

Conceptual blinders can be terrible, making it very hard to act against current interests to achieve even greater future interests. So people will need help with that, both those who need to change and those who help them. The crux of the problem seems to be how powerful concepts, disconnected from their contexts, so simplify what they show how to do they become inescapable, neatly hiding all the side effects too. What will most help people out of those traps is probably different in every case, but shared experience, humor, and irony seem higher on the list than promises and explanations. It is a matter of motivating internal change, often not responsive to external pressures unless everyone feels them at once.

# 4 Acknowledgments

During the pandemic, I have enjoyed a great deal of uninterrupted time to do the work, good health, a secure home, a good roommate, and adequate income, which together have been a godsend. What most contributed and made the work possible in the first place was having some parts of this fresh question to listen for answers to, constantly refreshed, and inspired by insights and reactions from many others over the years. So, I dedicate the work to all the many people who have shared their visions.

# 5 Data Sources

## 5.1 Climate

1. Scripps Global average Atmospheric CO2 ppm, combining splined ice core data before 1958, and yearly average mountain top measurements from of Mauna Loa and Antarctica including 1958 thereafter.
   http://scrippsco2.ucsd.edu/data/atmospheric_co2/icecore_merged_products

## 5.2 Economy

2. GDP (PPP) 1971 – 2016*     Fig 8
   Archived IEA PPP data extended with recent World Bank data, see Fig 13 for illustration
   WB:
   https://data.worldbank.org/indicator/NY.GDP.MKTP.PP.CD?end=2016&start=1990

3. World economic energy use 1965-2017 –   Fig 8
   BP: https://www.bp.com/en/global/corporate/energy-economics/statistical-review-of-world-energy/downloads.html

4. Modern CO2 Emissions – 1971-2016,     Fig 8
   Archived IEA CO2 data extended with WRI CO2 emissions:https://www.wri.org/resources/data-sets/cait-historical-emissions-data-countries-us-states-unfccc

5. Historical Co2 Emissions 1751-2013     Fig 8
   US DOE DOE CDIAC data: https://cdiac.ess-dive.lbl.gov/ftp/ndp030/global.1751_2014

6. World Meat Production – 1961-2016 Fig 8
   Rosner - OurWorldInData: https://ourworldindata.org/meat-and-seafood-production-consumption

7. World Food Production – 1961-2016 Fig 8
   FAO: http://www.fao.org/faostat/en/#data/QI

Holistic Natural Systems - Design & Steering# 6   References

## 6.1 Image References

<u>Fig 5 ref – Figures of all ages of the normal lifecycle.</u>

Guzaliia Filimonova / Getty Images fair use exception request. Also used by **Richard Nordquist** Updated on July 17, 2019 in his [Semantic Field Definition](#). [Getty agreement](#) Thank you for your inquiry, we are tracking it under the reference number 03062829 for you. https://www.gettyimages.com/customer-support

## 6.2 Text References

Bäck, A. (2006). The concept of abstraction. *The Society for Ancient Greek Philosophy Newsletter*. https://orb.binghamton.edu/cgi/viewcontent.cgi?article=1375&context=sagp

Bateson, N. (2017). Warm data: Contextual research and the evolution of science. Transdisciplinary Studies on Culture (i) Education No. 12, 35-40. *Scientific Yearbook of the University of Kujawsko-Pomorska in Bydgoszcz*. https://tinyurl.com/jt7896ve

Bell, B. (1971). The dark ages in ancient history. I. The first dark age in Egypt. *American Journal of Archaeology*, 75(1), 1-26. https://doi.org/10.2307/503678

Boulding, K. E. (1953). Toward a general theory of growth. *The Canadian Journal of Economics and Political Science / Revue Canadienne d'Economique et de Science Politique*, 19(3), 326–340. https://www.jstor.org/stable/138345?seq=1

Burkert, W. (1985). *Greek religion*. Harvard University Press.

Cabrera, D., Colosi, L., & Lobdell, C. (2008). Systems thinking. *Evaluation and program planning*, 31(3), 299-310. https://ecommons.cornell.edu/bitstream/handle/1813/2860/DerekCabreraDissertation.pdf

Cabrera, D., & Cabrera, L. (2022). DSRP Theory: A Primer. *Systems*, 10(2), 26. https://www.mdpi.com/2079-8954/10/2/26/pdf

Chew, S. C. (2007). *The recurring dark ages: ecological stress, climate changes, and system transformation* (Vol. 2). Rowman Altamira.J Henshaw                                              48                                              14-Aug-22

Holistic Natural Systems - Design & Steering

Holistic Natural Systems - Design & Steering

– End –